\documentstyle[12pt]{article}

\newcommand{\Ld}{\Lambda}

\newcommand{\pa}{\partial}
\newcommand{\co}{C_{\frac{5}{2} \frac{5}{2}}^{4}}

\newcommand{\str}{\rule[-2.5mm]{0mm}{7mm}}

\newcommand{\extraspace}{\addtolength{\abovedisplayskip}{2mm}
                        \addtolength{\belowdisplayskip}{2mm}
                        \addtolength{\abovedisplayshortskip}{2mm}
                        \addtolength{\belowdisplayshortskip}{2mm}}
\newcommand{\be}{\begin{equation}\extraspace}

\newcommand{\ee}{\end{equation}}

\newcommand{\bea}{\begin{eqnarray}\extraspace}
\newcommand{\beastar}{\begin{eqnarray*}\extraspace}
\newcommand{\eea}{\end{eqnarray}}
\newcommand{\eeastar}{\end{eqnarray*}}
\newcommand{\nonu}{\nonumber \\[2mm]}
\newcommand{\np}{Nucl.Phys.\ }

\newcommand{\cmp}{Comm.Math.Phys.\ }
\newcommand{\pl}{Phys.Lett.\ }

\begin{document}

\begin{flushright}
                                     ITP.SB-91-61 \\ Nov.'91
\end{flushright}

\vspace{5mm}

\begin{center}
{\LARGE $c=5/2$ Free Fermion Model of $ {WB_{2}}$
Algebra}\\\vspace{1cm}
{\large Changhyun Ahn\footnote{email: AHN@MAX.PHYSICS.SUNYSB.EDU}} \\
\vspace{7mm}
\str {\em Institute for Theoretical Physics} \\
\str {\em State University of New York at Stony Brook} \\
\str {\em Stony Brook, NY 11794-3840} \\
\vspace{2cm}
\end{center}

We investigate the explicit construction of the $WB_{2}$ algebra, which
is closed and associative for all values of the central charge
$c$, using the Jacobi identity and show the agreement with the
results studied previously.
Then we illustrate a realization of $c=\frac{5}{2}$ free
fermion model, which is $m \rightarrow \infty$ limit of unitary
minimal series, $c ( WB_{2} )=\frac{5}{2} (1-\frac{12}{
(m+3)(m+4)
})$ based on the cosets $( \hat{B_{2}} \oplus \hat{B_{2}},
\hat{B_{2}
})$ at level $(1,m).$ We confirm by explicit computations that
the bosonic currents in the $ WB_{2}$ algebra are indeed given
by the Casimir operators of  $\hat{B_{2}}$ .

\vspace{3mm}

\baselineskip=18pt

\newpage


\baselineskip=18pt

\section{Introduction}

\setcounter{equation}{0}

\indent

Recently much progress has been made in our understanding the
structure of two-dimensional rational conformal field theories
(RCFTs) for which the spectrum contains a finite number of irreducible
representations of chiral algebra, the operator product algebra
(OPA) of all the holomorphic fields in the theory. Cardy \cite{cardy}
was able to show that the sum over irreducible representations
of the conformal algebra must be infinite for the central charge
$c \geq 1$. Therefore in order to study RCFTs  with $c \geq 1$
,  extended conformal algebras which
consist of the Virasoro algebra and additional higher spin currents,
 play an important
role. Large classes of  extended Virasoro algebras are known:
affine Kac-Moody algebra \cite{kz}, superconformal algebra ($WB_{1
}$ ) \cite{gko}, and a class of $W$ algebras. Some examples of $W$
algebras have been established: $W_{3}$ algebra \cite{zamo}, $
W_{n}$ algebra \cite{fl}, and supersymmetric $W$ algebra
\cite{swa}.

Fateev and Lukyanov \cite{fl2} have considered an infinite dimensional,
associative quantum algebra $WB_{n}$, corresponding to the Lie
algebras $B_{n}$, consisting of $(n+1)$ fields of spin $ 4, 6,
\cdots, 2n$, and a fermion field of spin $(n+1/2)$ in addition to the
spin  $2$ energy momentum tensor. The addition of fermion
field was necessary for the quantization of the Hamiltonian
structures associated with Lie algebra whose simple roots have
a different length. They conjectured that these $(n+1)$ fields
form a closed OPA. These constructions produce representations of the
algebras over a range of values of $c$. For $n=1$, the $WB_{1}$ algebra
coincides with the $N=1$ super Virasoro algebra.

On the other hand, Watts \cite{watts} has shown that $W$ structures
may be constructed
in coset models of the form $( \hat {g_{n}} \oplus \hat {g_{n}},\hat {
g_{n}} )$ at level $( 1, m)$ with $g_{n}$ one of the finite Lie algebras
ABDE and for $m$ sufficiently large. An explicit expression for a fermion
field which commutes with the diagonal subalgebra was given in terms of
currents of $ \hat{B_{n}}$. The discrete series of $c$ \cite{fl2}
coincide with values obtained from coset models based on $\hat{B_{n}}$
algebra.
It is expected, but has not been established that
 his construction
are related to those in the free field construction \cite{fl2}.
Recently, he proved that the symmetry of Lie superalgebras of
$B(0, n)$ Toda theory was given
by the classical Poisson bracket analogue of the $WB_{n}$ algebra
\cite{watts2}.

It wasn't proven that  algebra $WB_{2}$  are associative for all
values of $c$ until Figueroa-O'Farrill et al. showed that the existence
of $WB_{2}$ explicitly using the perturbative conformal bootstrap
\cite{fst}. They have been able to show the equivalence with the
findings conjectured by
Fateev and Lukyanov \cite{fl2} using Coulomb gas realization.
We make an explicit construction of the Casimir of $WB_{2}$ from a
fermion proposed by Watts in \cite{watts}.

This paper is organized as follows. In section 2, we look at the
explicit construction of $WB_{2}$ algebra using the associativity of
Laurent expansion operators, which we investigate by considering a
graded Jacobi identity for Laurent expansion modes.
In section 3, we would like to understand how $WB_{2}$ currents
emerge from fields of $\hat{B_{2}}$ algebra and prove a realization of
$c=5/2$ representation of section 2.
 Finally  section 4 contains a few concluding remarks.

\section{$\bf WB_{2}$ Algebra}

\setcounter{equation}{0}

\indent

As our starting point, we follow the analysis in \cite{fl2}.
The presence of $W_{
2}$ current of dimension $2$ and $W_{4}$ current of dimension $4$ can
be
understood from the operator product expansion (OPE) of $d$ current
of dimension $5/2$ with itself:
\bea
\frac{30}{(2c+25)} d(z) d(w) & = & \frac{1}{(z-w)^{5}} \frac{2}{5} c+
                                   \frac{1}{(z-w)^{3}} 2 W_{2} (w)+
\frac{1}{(z-w)^{2}} \pa W_{2} (w)\nonu
                             &   & +\frac{1}{(z-w)} [\frac{60}{(2c+25)}
W_{4}
                                   (w)+\frac{1}{2} \pa^{2} W_{2}(w) ]
+\cdots .
\eea
Notice that $W_{4}$ is {\em not} a primary field under
the $W_{2} (z)= T(z)$
energy momentum tensor which generates the Virasoro algebra with a central
charge $c$. The above OPE can be rewritten as the more familiar form
in
the following way :
\bea
& & U(z) U(w)  =  \frac{1}{(z-w)^{5}} \frac{2}{5} c +\frac{1}
{(z-w)^{3}} 2T(w)+\frac{1}{(z-w)^{2}} \pa T(w) \nonu
& & +\frac{1}{(z-w)} [\frac{3}{10} \pa ^{2} T(w) +\frac{27}
   {(5c+22)} \Lambda(w)+ C_{\frac{5}{2} \frac{5}{2}}^{4} V (w) ]
+\cdots,
\eea
where
\bea
 \sqrt{\frac{30}{(2c+25)}} d (z) = U(z), \; \Lambda (z) =
T^{2} (z)-\frac{3}{10} \pa^{2} T (z),
\eea
and
\bea
 W_{4} (z)  =  -\frac{(2c+25)^{2}}{120 (5c+22)} \pa^{2} T (z)+\frac{
                 9(2c+25)}{20 (5c+22)} T^{2} (z)
            +\frac{(2c+25)}{60} \co V (z).
\eea

It is convenient to use $V(z)$, which is a Virasoro primary field
of dimension 4 , rather
than $W_{4}(z)$ for the future developments of $WB_{2}$ algebra. $\co$
is a
coupling constant and to be determined later.
Extended conformal algebra
by a single primary field of $5/2$ conformal dimension was found in
\cite{zamo}.
Here enlarging the algebra by additional spin $4$ field leads to the
appearance of $V(z)$ term of eq. $(2.2).$ We would expect that
$\co$ should vanish for $c=-13/14$.
On the other hand,
it has been proven that one extra spin $4$ current algebra was
determined by the closure of Jacobi identity \cite{peter,ht}. We
 are dealing with two-dimensional RCFTs,
 in which there exist conserved currents $U(z)$ of
spin $
5/2$ and $V(z)$ of spin $4$, and would expect that $U(z)$ field
dependence
should appear in the OPE $V(z) V(w)$. Recall that $U^{2}(z)$
can be written as the combinations of $\pa^{3} T(z), \pa \Ld (z)
\; \mbox{and} \; \pa V (z)$ from eq. $(2.2)$. We can see that the
nontrivial $U(z)$ field dependence has the form of $\pa U U(z)$
( or $U \pa U(z)$ ).
The next step is to work out
the
OPE $V(z) V(w)$. Through an analysis of \cite{ht} with the symmetry
under the interchange of $z$ and $w$, it leads to the
following
results:
\bea
& & V(z) V(w)  =  \frac{1}{(z-w)^{8}} \frac{c}{4} +A [\frac{1}{(z-w)
^{6}}
                2 T(w)+\frac{1}{(z-w)^{5}} \pa T(w)\nonu
& & -\frac{1}{(z-w)^{3}} \frac{1}{12} \pa^{3} T(w)+\frac{1}
   {(z-w)} \frac{1}{120} \pa^{5} T(w) ]+B [\frac{1}{(z-w)
                ^{4}} 2 \pa^{2} T(w)\nonu
& & +\frac{1}{(z-w)^{3}} \pa^{3} T(w)-\frac{1}{(z-w)}
                \frac{1}{12} \pa^{5}T(w)]+C[\frac{1}{(z-w)^{2}} 2
                \pa^{4} T(w)\nonu
& & +\frac{1}{(z-w)} \pa^{5} T(w)]+D [\frac{1}{(z-w)^{4}}
                2 \Ld (w)+\frac{1}{(z-w)^{3}} \pa \Ld (w)\nonu
& & -\frac{1}{(z-w)} \frac{1}{12} \pa^{3} \Ld (w)]+E[\frac{1}
                {(z-w)^{2}} 2 \pa^{2} \Ld (w)+\frac{1}{(z-w)} \pa^{3}
                \Ld (w)]\nonu
& & +F[\frac{1}{(z-w)^{2}} 2 \Xi (w)+\frac{1}{(z-w)} \pa
                \Xi (w)]+G[\frac{1}{(z-w)^{2}} 2 \Delta (w)\nonu
& & +\frac{1}{(z-w)} \pa \Delta (w)]+H[\frac{1}{(z-w)^{4}}
                2 V(w)+\frac{1}{(z-w)^{3}} \pa V(w)\nonu
& & -\frac{1}{(z-w)} \frac{1}{12} \pa^{3} V(w)]+I[\frac{1}
                {(z-w)^{2}} 2 \pa^{2} V(w) +\frac{1}{(z-w)} \pa^{3}
V(w)]
                \nonu
& & +J[\frac{1}{(z-w)^{2}} 2 \Omega (w) +\frac{1}{(z-w)} \pa
                \Omega (w)]\nonu
& & +K[\frac{1}{(z-w)^{2}} 2 \Gamma (w)+\frac{1}{(z-w)} \pa
                \Gamma (w)]+\cdots,
\eea
with
\bea
& & \Xi (z)  =  \pa T \pa T(z) -\frac{3}{70} \pa^{4} T(z)-\frac{31}
                {7(5c+ 22)} \pa^{2} \Ld (z),\nonu
& & \Delta (z)=T \Ld (z)-\frac{3}{10} \pa^{2} \Ld (z), \; \Omega
               (z)=T V(z),\nonu
& & \Gamma (z)=\pa U U(z)-\frac{5}{18} \pa T \pa T (z).
\eea

Note that the operator $\Gamma (z)$ defined as above has no central
term which makes the calculation easier. To guarantee that the
complete OPA be associative, $K$ terms are inevitable. In order to
fix the coefficients in the OPE we consider the condition of
associativity. We will use the Jacobi identities for
 Laurent expansion modes. The commutation
relation $[V_{m}, V_{n}]$ can be obtained by contour integral, as
usual. For the commutators of newly defined operators of $(2.6)$ with
Virasoro
operators $L_{m}$, see the reference \cite{ht}. Also we need to know $
[ L_{m}, \Gamma_{n}]=\oint_{C_{0}} (dw/2\pi i) w^{n+5} \oint_{C_{w}}
(dz/2\pi i) z^{m+1} T(z) \Gamma (w)$ :
\bea
& & [L_{m},\Gamma_{n}]  =  \frac{m(m^{2}-1)}{5! 180}[2020(m-2)(m-3)
                         -100(-10c+27)(m-2)\nonu
& & \times (m+n+2)+2020(m+n+2)(m+n+3)] L_{m+n}\nonu
& & +\frac{m}{3! 18 (5c+22)}[3159(m^{2}-1)-3(m+1)(-100c+775)
   (m+n+4)]\Ld_{m+n}\nonu
& &+\frac{m}{12} \co [13(m^{2}-1) -15(m+1)(m+n+4)]V_{m+n}
+(5m-n)\Gamma_{m+n}.
\eea
After a straightforward computation of the following Jacobi identity
\bea
[L_{m}, [V_{n}, V_{p}]]+[V_{p},[L_{m},V_{n}]]+[V_{n},[V_{p},L_{m}]]=0,
\eea
we can have the intermediate results:
\bea
& & A=1, \; B=\frac{3}{20}, \; C=\frac{1}{168}, \; D=\frac{21}
{(5c+22)}\nonu
& & E=\frac{22}{7(5c+22)}-\frac{101}{36(5c+22)}K,\nonu
& & F=-\frac{3(19c-524)}{20(2c-1)(7c+68)}+\frac{7(50c^{2}+507c-968)}
{90(2c-1)(7c+68)}K\nonu
& & G=\frac{12(72c+13)}{(2c-1)(7c+68)(5c+22)}-\frac{(734c+49)}{(2c-1)(
7c+68)(5c+22)}K\nonu
& & H=\frac{3(c+24)}{28}J+\frac{19}{28} \co K, \; I=\frac{(5c+64)}
{336} J-
\frac{5}{112} \co K.
\eea

We have still now three free parameters, $\co, J, \mbox{and} \;K$. As
you can
see, the above results reduce to ones in \cite{ht,peter} when $K$ goes to zero.
We can check that the other Jacobi identity involving $L_{m}, U_{n},
\mbox{and} \;U_{p}$ doesn't give any further restriction on these free
parameters.

Now we investigate the OPE $V(z) U(w)$ in order to obtain the complete
algebra. Then some local operators appearing in the r.h.s. of $V(z)
U(w)$ are made of the products $T(w)$, $U(w)$ and their derivatives.
The most general form of the OPE of $V(z) U(w)$ is
\bea
& & V(z) U(w)  =  \frac{a}{(z-w)^{4}} U(w)+\frac{b}{(z-w)^{3}} \pa
U(w)                          \nonu
& & +\frac{1}{(z-w)^{2}}[d \pa^{2} U(w)+e T U (w)]\nonu
& & +\frac{1}{(z-w)}[f \pa^{3} U(w)+g \pa T U(w)+h \pa (
                T U)(w)] +\cdots.
\eea
In a similar manner, the unknown structure constants are determined by
the following Jacobi identity :
\bea
[L_{m}, [V_{n}, V_{p}]]+[U_{p}, [L_{m}, V_{n}]]+[V_{n}, [U_{p}, L_{m}
]]=0.
\eea
Therefore it leads to the following results,
\bea
& & \co a=\frac{15(14c+13)}{4(5c+22)}, \; \co b=\frac{3(14c+13)}
{(5c+22)}\nonu
& & \co d=\frac{5(c+8)(14c+13)}{2(5c+22)(2c+25)}, \; \co e=\frac{45
(14c+13)}{
(5c+22)(2c+25)}\nonu
& & \co f=\frac{(2c-5)}{(2c+25)}, \; \co g=-\frac{162c+2025}
{(5c+22)(2c+25)}\nonu
& & \co h=\frac{6(82c+215)}{(2c+25)(5c+22)}.
\eea
In order to get $\co, J, \mbox{and} \;K$ completely, we should use
explicit
construction of $[V_{m},\Ld_{n}], \{U_{m}, (T U)_{n}\},\;\mbox{and}
\; \{U_{m}, (\pa T U)_{n}\}$ to satisfy the following Jacobi identity:
\bea
[V_{m},\{U_{n}, U_{p}\}]-\{U_{p}, [V_{m}, U_{n}]\}-\{U_{n},[V_{m},
U_{p}]\}
=0.
\eea
But we refrain from giving the explicit expressions for the above
( anti)
commutators, which are rather complicated and not illuminating.
 Finally, with
\bea
& & \co =\sqrt{\frac{6(14c+13)}{(5c+22)}}, \; J=\frac{4\sqrt{6}
(7c-115)}{
(2c+25)\sqrt{(5c+22)(14c+13)}}\nonu
& & K=\frac{30(5c+22)}{(2c+25)(14c+13)},
\eea
putting all together, all the coefficients can be determined in terms
of {\em only} the Virasoro central charge $c$:
\bea
& & a=\frac{15\sqrt{(14c+13)}}{4\sqrt{6(5c+22)}}, \; \; b=\frac{3
\sqrt{14c+
13}}{\sqrt{6(5c+22)}}\nonu
& & d=\frac{5(c+8) \sqrt{14c+13}}{2(2c+25)\sqrt{6(5c+22)}}, \;\;
e=\frac{
45\sqrt{14c+13}}{(2c+25)\sqrt{6(5c+22)}}\nonu
& & f=\frac{(2c-5)\sqrt{5c+22}}{(2c+25)\sqrt{6(14c+13)}}, \;\;
g=-\frac{
162c+2025}{(2c+25)\sqrt{6(5c+22)(14c+13)}}\nonu
& & h=\frac{6(82c+215)}{(2c+25)\sqrt{6(5c+22)(14c+13)}},\nonu
& & E=\frac{3696c^{2}+31957c-34870}{42(2c+25)(5c+22)(14c+13)},
\;\; F=
\frac{2158c+21305}{60(2c+25)(14c+13)}\nonu
& & G=\frac{54(32c-5)}{(2c+25)(5c+22)(14c+13)}, \;\; H=\frac{3
\sqrt{6}(
2c^{2}+83c-490)}{(2c+25)\sqrt{(5c+22)(14c+13)}}\nonu
& & I=\frac{\sqrt{6} (10c^{2}-197c-2810)}{24(2c+25)\sqrt{(5c+22)(14c+
13)}}.
\eea

Note that $WB_{2}$ algebra is a subalgebra of Zamolodchikov's spin
$5/2$ algebra \cite{zamo} for $c=-13/14$ because $\co$ vanishes for
this
value of $c$. These results are in agreement with the findings of
\cite{fst}.\footnote{ In this way we discovered that there is a misprint
of eq.(13) of ref.\cite{fst}. The factor 5 in the denominator should be
in the numerator.}  The one thing which we would like to stress
is the fact that all the remaining Jacobi identities, $ [ U_{m},
[V_{n}, V_{p}]]+\mbox{cycl}.=0, [U_{m},\{ U_{n},U_{p}\}]+\mbox{cycl}
.=0,
\;\mbox{and}\;
[V_{m},[V_{n},V_{p}]]+\mbox{cycl}.=0 $, are consistent with the above
results after a long calculation with $ {\em Mathematica^{TM}\/}$ \cite
{mt}.

\section{The Five Free Fermion Model}

\setcounter{equation}{0}

\indent

The coset models \cite{watts} are defined in terms of the currents
$E_{(1)}^{ab}(z),
$ and $E_{(2)}^{ab}(z)$, of level $1$ and $m$, respectively, which
generate the algebra $ g=\hat{B_{2}} \oplus \hat{B_{2}}$. The generator
of the diagonal subalgebra $g^{\prime}=\hat{B_{2}}$, which has level $
m^{\prime}=1+m$, is given by
\be
{E^{\prime}}^{ab}(z)=E_{(1)}^{ab}(z)+E_{(2)}^{ab}(z).
\ee
The coset Virasoro algebra is generated by the difference $T_{(1)}(z)+
T_{(2)}(z)-T^{\prime}(z),$ where $T_{(1)}(z)$ and $T_{(2)}(z)$ are
Sugawara stress energy tensors:
\be
\tilde{T}(z)=-\frac{1}{16} E_{(1)}^{ab} E_{(1)}^{ab}(z)-\frac{1}{4(m+3)
} E_{(2)}^{ab} E_{(2)}^{ab}(z)+\frac{1}{4(m+4)} {E^{\prime}}^{ab}
{E^{\prime}}^{ab}(z).
\ee
Of course, $\tilde{T}(z)$ commutes with ${E^{\prime}}^{ab}(z)$.
The coset central charge of the unitary minimal models for $WB_{2}$
is
\be
\tilde{c}=c(WB_{2})=\frac{5}{2}+\frac{10m}{m+3}-\frac{10(m+1)}{m+4}
=\frac{5}{2} (1-\frac{12}{(m+3)(m+4)})
\ee
where $  m=1,2, \cdots .$

In this section we focus on the limit $m\rightarrow \infty$, which
gives us a model
of $c=5/2$ that is invariant under the affine Lie algebra $\hat {B
_{2}}$ at level $1$. This model can be represented by $5$ free
fermions
$\psi^{a}$ of dimension $1/2$, where the index $a$ takes values in
the
adjoint representation of $B_{2}$ and $a=1, \cdots ,5$. We will show
how the currents of  $WB_{2}$ can be constructed from these free
fermion fields or basic fields $E^{ab}$ of $\hat{B_{2}}$
and consider their OPA.

The defining OPE of the basic fermion fields is given by, as usual,
\be
\psi^{a} (z) \psi^{b} (w)=\frac{1}{(z-w)} \delta^{ab} +\cdots.
\ee
We can define dimension $1$ currents $E^{ab} (z)$ as composites of the
free fermions
\be
E^{ab} (z)=\psi^{a} \psi^{b} (z)
\ee
which satisfy, at level $1$, the usual $\hat{B_{2}}$ OPE
\bea
& & E^{ab}(z) E^{cd}(w)  =  \frac{1}{(z-w)^{2}}(\delta^{bc} \delta^
{ad}-
                          \delta^{ac} \delta^{bd})\nonu
& & +\frac{1}{(z-w)}[\delta^{bc} E^{ad}(w)+
                          \delta^{ad} E^{bc}(w)-\delta^{ac} E^{bd}(w)
 -\delta^{bd} E^{ac}(w)]+\cdots.
\eea
Watts \cite{watts} has pointed out that $ U(z) $ of dimension $5/2$,
which is
invariant under the horizontal subalgebra, can be expressed as
follows
using $B_{2}$ invariant $\epsilon^{abcde}$ tensor.
\be
U(z)=\frac{1}{120} \epsilon^{abcde} \psi^{a} E^{bc} E^{de}(z).
\footnote{Multiple composite operators are always regularized
from the right to left, unless otherwise stated. The normalization of
U(z) is chosen such that $\epsilon^{abcde} \epsilon^{abcde}=120$.}
\ee
After a tedious calculation, using the rearragement lemmas \cite{bais}
, we
arrive at the following result for the OPE of $U(z)$ with $U(w)$
\footnote{A product of two of $\epsilon $ tensors can be expressed
as a determinant in which the entries are $\delta$'s.},
\bea
U(z) U(w) & = & \frac{1}{(z-w)^{5}}-\frac{1}{(z-w)^{3}} \psi^{a} \pa
                \psi^{a} (w)-\frac{1}{(z-w)^{2}} \frac{1}{2}
               \psi^{a} \pa^{2} \psi^{a} (w)\nonu
          &   & +\frac{1}{(z-w)} [\frac{1}{2} \psi^{a} \pa \psi^{a}
                \psi^{b} \pa \psi^{b}(w)-\frac{1}{6} \psi^{a} \pa^{3
                } \psi^{a} (w)] +\cdots.
\eea
During this calculation, we used the fact that $\epsilon^{abcde}
\epsilon^{afghi} (( E^{bc} E^{de} )( E^{gh} E^{ij} ))(z)=864 \psi
^{a}\pa \psi^{a} \psi^{b} \pa \psi^{b} (z)-384 \psi^{a} \pa^{3} \psi
^{a}(z)$.
Comparing this with eq. $(2.2)$, one can readily see that
\bea
V(z) & = & \frac{1}{8\sqrt{69}}[7\psi^{a} \pa \psi^{a} \psi^{b} \pa
           \psi^{b}(z)-6\pa \psi^{a} \pa^{2} \psi^{a}(z)+\frac{2}{3}
           \psi^{a} \pa^{3} \psi^{a}(z)]\nonu
     & = & \sqrt{\frac{23}{192}} [\frac{28}{23} T^{2}(z)+\frac{33}{
           23} \pa^{2} T (z)+\psi^{a} \pa ^{3} \psi^{a} (z)]\nonu
     & = & -\frac{1}{40 \sqrt{69}} E^{ab} E^{cd} E^{ac} E^{bd}(z)
\eea
which is the unique ( up to a normalization ) dimension $4$ primary
field under the energy momentum tensor, $T(z)=-\frac{1}{2} \psi^{
a} \pa \psi^{a} (z)=-\frac{1}{16} E^{ab} E^{ab}(z)$ which is the form of
 the second order Casimir. For $B_{2}$ algebra, the number of independent
Casimirs equals the rank of $B_{2}$ (=2). Therefore we have in addition
to the
second Casimir only a fourth order Casimir given by $ (3.9)$.
The fact that the fourth order Casimir operator is generated in the
OPE $U(z) U(w)$ confirms Casimir algebras consisting of $ T(z)
\;\mbox{and} \;V(z)$ are {\em not} the usual spin-4 algebras
\cite{ht,peter}.

 In order to find
the complete structure of
$WB_{2}$, one has to take the OPE $U(z) V(w)$ :
\bea
& & \sqrt{\frac{192}{23}} U(z) V(w)  =  \frac{1}{(z-w)^{4}}
\frac{120}{
23} \frac{1}{120} \epsilon^{abcde} \psi^{a} E^{bc} E^{de}(w)
\nonu
& & +\frac{1}{(z-w)^{3}} \frac{24}{
23} \frac{1}{120} \epsilon^{ab
cde} \pa ( \psi^{a} E^{bc} E^{de} )(w)\nonu
& &+\frac{1}{(z-w)^{2}} \frac{1}{
120}[\frac{48}{23} \frac{15}{4} \epsilon^{abcde} \psi^{a} \psi^{b}
\psi^{c} \psi^{d} \pa^{2} \psi^{e}(w)-\frac{8}{23} \epsilon^{abcde}
\pa^{2} (\psi^{a} E^{bc} E^{de})(w)]\nonu
& &+\frac{1}{(z-w)} \frac{1}{120}[
-\frac{8}{23} \frac{15}{4} \epsilon^{abcde} \pa (\psi^{a} \psi^{b}
\psi^{c} \psi^{d} \pa^{2} \psi^{e})(w)\nonu
& & +\frac{54}{23} \{ \frac{2}{3}
\epsilon^{abcde} \pa^{3} (\psi^{a} E^{bc} E^{de})(w)+5 \epsilon^{
abcde} \pa^{3} \psi^{a} E^{bc} E^{de}(w)\nonu
& & -\frac{5}{3} \epsilon^{abcde}
\psi^{a} \pa^{3} (E^{bc} E^{de})(w)+20 \epsilon^{abcde} \psi^{a}
\psi^{b} \psi^{c} \pa \psi^{d} \pa^{2} \psi^{e}(w) \}]\nonu
& & =  \frac{1}{(z-w)^{4}} \frac{120
}{23} U(w)+\frac{1}{(z-w)^{3}} \frac{24}{23} \pa U(w) +\frac{1}
{(z-w)^{2}}[\frac
{48}{23} TU(w)\nonu
& & -\frac{8}{23} \pa^{2} U(w)]
 +\frac{1}{(z-w)}[-\frac{8}{23}
 \pa (TU)(w)+\frac{54}{23} \pa TU(w)]
 +\cdots.
\eea
Basically, it agrees with the expressions given in $(2.10)$ and $(2.15)$
for $c=\frac{5}{2}$.
We are ready to consider the OPE of $V(z) V(w)$. We explicitly
computed $V(z) V(w)$, obtained from eq. $(3.9)$, which is given by
\bea
& & V(z) V(w)  =  \frac{1}{(z-w)^{8}} \frac{5}{8} +\frac{1}{(z-w)^
{6}}
2 T(w) +\frac{1}{(z-w)^5} \pa T(w)\nonu
& & +\frac{1}{(z-w)^{4}} 2 [\frac{3}{20} \pa^{2} T(w)+
\frac{14}{23} \Ld(w)-\frac{27}{4\sqrt{69}} V(w)]\nonu
& & +\frac{1}{(z-w)^{3}}[\frac{1}{15} \pa^{3} T(w)+\frac{
14}{23} \pa \Ld (w)-\frac{27}{4\sqrt{69}} \pa V(w)]\nonu
& & +\frac{1}{(z-w)^{2}} 2[\frac{1}{168} \pa^{4} T(w)+
\frac{9083}{278208} \pa^{2} \Ld (w)+\frac{15}{184} \Delta (w)\nonu
& & +\frac{89}{288} \Xi(w)-\frac{9}{4\sqrt{69}} \pa^{2}
V(w)-\frac{13}{2\sqrt{69}} \Omega (w)+\frac{23}{32} \Gamma(w)]\nonu
& & +\frac{1}{(z-w)}[\frac{1}{560} \pa^{5} T(w)-\frac{
5029}{278208} \pa^{3} \Ld(w)+\frac{15}{184} \pa \Delta(w)\nonu
& &+\frac{89}{288} \pa \Xi(w)-\frac{27}{16\sqrt{69}} \pa
^{3} V(w)-\frac{13}{2\sqrt{69}} \pa \Omega (w)+\frac{23}{32} \pa
\Gamma(w)]
\eea
where $\Ld(z), \Delta(z), \Xi(z), \Omega(z) \;\mbox{and}\; \Gamma(z)$
which are expressed in terms of $\psi^{a}(z)$'s according to
\bea
& & \Ld (z)=\frac{1}{4} \psi^{a} \pa \psi^{a} \psi^{b} \pa \psi^{b}
(z)+
\frac{21}{40} \pa \psi^{a} \pa^{2} \psi^{a} (z)-\frac{7}{120} \psi^{
a} \pa^{3} \psi^{a}(z)\nonu
& & \Delta(z) =  -\frac{1}{8} \psi^{a} \pa \psi^{a} \psi^{b} \pa
\psi^{b}\psi^{c} \pa \psi^{c} (z)-\frac{3}{20} \psi^{a} \pa^{2}
\psi^{a}\psi^{b} \pa^{2} \psi^{b} (z)\nonu
& & -\frac{63}{80} \psi^{a} \pa \psi^{a} \pa \psi^{b} \pa^
{2} \psi^{b} (z)+\frac{7}{80} \psi^{a} \pa \psi^{a} \psi^{b} \pa^{3}
 \psi^{b} (z)-\frac{161}{400} \pa^{2}  \psi^{a} \pa^{3} \psi^{a} (z)
\nonu
& & +\frac{49}{1600} \pa \psi^{a} \pa^{4} \psi^{a} (z)+
\frac{7}{1600}\psi^{a} \pa^{5} \psi^{a} (z)\nonu
& & \Xi (z)  =  -\frac{31}{483} \psi^{a} \pa \psi^{a} \pa \psi^{b}
\pa^{2} \psi^{b} (z)-\frac{31}{483} \psi^{a} \pa \psi^{a} \psi^{b
} \pa^{3} \psi^{b}(z)\nonu
& & +\frac{359}{1932} \psi^{a} \pa^{2} \psi^{a} \psi^{b}
\pa^{2} \psi^{b}(z)+\frac{1084}{7245} \pa^{2} \psi^{a} \pa^{3} \psi
^{a} (z)\nonu
& &-\frac{32}{7245} \psi^{a} \pa^{5} \psi^{a} (z)+\frac{
86}{7245} \pa \psi^{a} \pa^{4} \psi^{a} (z)\nonu
& &\Omega(z)  =  \sqrt{\frac{23}{192}} [-\frac{7}{46} \psi^{a} \pa
\psi^{a} \psi^{b} \pa \psi^{b} \psi^{c} \pa \psi^{c} (z)-\frac{
15}{46} \psi^{a} \pa \psi^{a} \pa \psi^{b} \pa^{2} \psi^{b} (z)
\nonu
& & +\frac{11}{46} \psi^{a} \pa \psi^{a} \psi^{b} \pa
^{3} \psi^{b} (z)+\frac{3}{23} \pa^{2} \psi^{a} \pa^{3} \psi^{a}
(z)\nonu
& & -\frac{7}{92} \pa \psi^{a} \pa^{4} \psi^{a} (z)+
\frac{3}{460} \psi^{a} \pa^{5} \psi^{a} (z)]\nonu
& &\Gamma (z)  =  -\frac{1}{6} \psi^{a} \pa \psi^{a} \psi^{b} \pa
\psi^{b} \psi^{c} \pa \psi^{c} (z) +\frac{1}{6} \psi^{a} \pa
\psi^{a} \psi^{b} \pa^{3} \psi^{b} (z)\nonu
& & +\frac{1}{18} \psi^{a} \pa^{2} \psi^{a} \psi^{b
} \pa ^{2} \psi^{b} (z) +\frac{1}{1080} \psi^{a} \pa^{5} \psi^{
a} (z)\nonu
& & -\frac{5}{108} \pa^{2} \psi^{a} \pa^{3} \psi^{a} (z),
\eea
which agree with the formulas $(2.6)$.

Crucial point to arrive at this result was to reexpress $\pa U U
(z)$ appearing in $1/(z-w)^2 $ of $V(z) V(w)$ in terms of 7
independent fields, consisting of composites of $\psi^{a}$ and their
derivatives, and recombine with $\pa^{2} V (z), T V (z), \mbox{and}
\; T (z)$ descendants. The OPE of two Virasoro primary fields can
be represented as the sum of Virasoro conformal families, i.e.,
Virasoro descendants and Virasoro primary fields \cite{bpz}. Then
we can identify unique Virasoro primary spin 6 field \cite {fst} with
\bea
& & \Phi (z)=\frac{1}{576}(5\psi^{a} \pa \psi^{
a} \psi^{b} \pa \psi^{b} \psi^{c} \pa \psi^{c} (z)-\frac{17}{4} \psi^{a} \pa
\psi^{a} \pa \psi^{b} \pa^{2} \psi^{b} (z)\nonu
& & -\frac{1}{5} \psi^{a} \pa^{2} \psi^
{a} \psi^{b} \pa^{2} \psi^{b} (z)+\frac{13}{20} \psi^{a} \pa \psi^{a}
\psi^{b} \pa^{3}
\psi^{b} (z)+\frac{1}{3} \pa^{2} \psi^{a} \pa^{3} \psi^{a} (z)\nonu
& & -\frac{1}{12} \pa \psi^{a}
\pa^{4} \psi^{a} (z)+\frac{1}{300} \psi^{a} \pa^{5} \psi^{a} (z)).
\eea
Of course, $ \Phi (z)$ is a descendant w.r.t. the full $ WB_{2}$
algebra.
Eq. (3.11) and (3.12) agree with the expression for the $WB_{2}$
algebra as given in
eq. $(2.5)$ and $(2.15) $ for $c=5/2$. The results obtained so far
can be summarized as follows.
In $c=5/2$ free fermion model, a consistent OPA can be made out of
energy momentum tensor $T(z) $ and additional currents $U(z) $ of
dimension $5/2$ and $V(z)$of dimension $4$, corresponding to a
fourth order Casimir of $B_{2}$.

\section{Conclusion}
\setcounter{equation}{0}
\indent

The remaining problem is to construct the $WB_{2}$ algebra in the coset
models
based on $(\hat{B_{2}} \oplus \hat{B_{2}}, \hat{B_{2}})$ at level
$(1, m)$, which can be   viewed as perturbations of the $m \rightarrow
\infty$ model discussed before. Then the dimension $5/2$ coset field
$\tilde{U}(z)$ was given in \cite{watts} where $\tilde{c} $ is as
in eq. $(3.3).$
 We can do calculate the
OPE $\tilde {U}(z) \tilde {U}(w)$. Then the dimension 4 coset field
$\tilde{V}(z)$ can be obtained from the singular part of OPE $
\tilde{U}(z) \tilde {U}(w)$. We would like to show explicitly that
the algebras, consisting of $\tilde{T}(z), \tilde{U}(z) \;\mbox{and}
\;\tilde{V}(z),$ closes
in the coset model.
Then,this construction will lead to an explicit realization of the $c <
5/2$ unitary representations of $ WB_{2}$ algebra. We leave it further
investigation \cite{work}.

\vspace{15mm}

It is a pleasure to thank M. Rocek and K. Schoutens for reading
the manuscript and
discussions, A. Sevrin for drawing my attention to
ref. \cite{fl2}.  This work was supported in part by grant NSF-91-08054.


\begin{thebibliography}{11}

\bibitem{cardy}
  J.L. Cardy, Operator Content of Two-Dimensional Conformally
  Invariant Theories, \np {\bf B270} [FS16] (1986) 186
\bibitem{kz}
  V.G. Knizhnik and A.B. Zamolodchikov, Current Algebra and
  Wess-Zumino Model in Two Dimensions, \np {\bf B247} (1984) 83
\bibitem{gko}
  P. Goddard, A. Kent and D. Olive, Virasoro Algebras and Coset
  Space Models, \pl {\bf B152} (1985) 88;
  Unitary Representations of the Virasoro and Super-Virasoro
  Algebra, \cmp {\bf 103} (1986) 105; D. Friedan, Z. Qiu and
  S. Shenker, Conformal Invariance, Unitarity and Critical
  Exponents in Two Dimensions, Phys.Rev.Lett. {\bf 52}(1984)
  1575
\bibitem{zamo}
  A.B. Zamolodchikov, Infinite Additional Symmetries in Two-
  Dimensional Conformal Quantum Field Theory, Theor.Math.Phys.
  {\bf 65} (1986) 1205

\bibitem{fl}
  V.A. Fateev and S.L. Lukyanov, The Models of Two-Dimensional
  Conformal Quantum Field Theory with $ Z_{n} $ Symmetry,
  Int.J.Mod.Phys. {\bf A3} (1988)
  507
\bibitem{swa}
  T. Inami, Y. Matsuo and I. Yamanaka, Extended Conformal Algebras
  With $N=1$ Supersymmetry, \pl {\bf 215B} (1988) 701;
  K. Hornfeck and E. Ragoucy, A Coset Construction for the Super
  $W_{3}$ Algebra, \np {\bf B340} (1990) 225; C. Ahn, K. Schoutens
  and A. Sevrin, The Full Structure of the Super $W_{3} $ Algebra,
  Int.J.Mod.Phys. {\bf A6} (1991) 3467; K. Schoutens and A. Sevrin,
  Minimal
  Super-$W_{N}$ Algebras in Coset Conformal Field Theories, \pl {\bf
  258B} (1991) 134; J.M. Figueroa-O'Farrill and S. Schrans,
  The Conformal Bootstrap and  Super $W$-Algebras, Leuven
  preprint KUL-TF-90/16; T. Inami, Y. Matsuo and I. Yamanaka, Extended
  Conformal Algebra with $ N=2$ Supersymmetry, Int.J.Mod.Phys. {\bf A5
  } (1990) 4441; H. Lu,
  C.N. Pope, L.J. Romans, X. Shen and X-J. Wang, Polyakov Construction of
  the $N=2$ Super-$W_{3}$ Algebra, \pl {\bf 264B} (1991) 91; T. Inami
  and H.
  Kanno, $N=2$ Super $W$ Algebras and Generalized $N=2$ Super KdV
  Hierarchies Based on Lie Superalgebras, YITP/K-928 May 1991; L.J. Romans,
  The $N=2$ Super-$W_{3}$ Algebra, USC-91/HEP06; K. Hornfeck,
  Supersymmetrizing the $W_{4}$ Algebra, \pl {\bf 252B} (1990) 357,
  Realizations for the $SW(7/2)$-Algebra and the Minimal Supersymmetric
  Extension of $WA_{3}$, King's College preprint, July 1991; D.H. Zhang
  and C.S. Huang, Spin $4$ and $7/2$ Extended Conformal Algebra, IC/90/211
\bibitem{fl2}
  S.L. Lukyanov and V.A. Fateev, Additional Symmetries and Exactly
  Soluble Models in Two-Dimensional Conformal Field Theory, Kiev,
  1988 ( preprints ITF-88-74R, 75R and 76R, Institute of Theoretical
  Physics, Ukranian Academy of Sciences )
\bibitem{watts}
  G.M.T. Watts, $WB$ Algebra Representation Theory, \np {\bf B339}
  (1990)  177 ; $W$- Algebras and Coset Models, \pl {\bf 245B}
  (1990) 65, DAMTP-90-37
\bibitem{watts2}
  G.M.T. Watts, $WB_{n}$ symmetry, Hamiltonian Reduction and $B(0,
  n)$ Toda Theory, \np {\bf B361} (1991) 311
\bibitem{fst}
  J.M. Figueroa-O'Farrill, S. Schrans and K. Thielmans, On the
  Casimir Algebra of $B_{2}$, \pl {\bf 263B} (1991) 378
\bibitem{ht}
  K. Hamada and M. Takao, Spin-4 Current Algebra, \pl {\bf 209B}
  (1988) 247 ; Erratum \pl { \bf 213B} (1988) 564
\bibitem{peter}
  P. Bouwknegt,
  Extended Conformal Algebras, \pl {\bf 207B} (1988) 295; D.H.
  Zhang, Spin-$4$ Extended Conformal Algebra, \pl {\bf 232B}
  (1989) 323
\bibitem{mt}
  S. Wolfram, ${\em Mathematica^{TM}\/}$, Addison-Wesley (1988)
\bibitem{bais}
  F.A. Bais, P. Bouwknegt, K. Schoutens and M. Surridge, Extensions
  of the Virasoro Algebra Constructed from Kac-Moody Algebras Using
  Higher Order Casimir Invariants; Coset Construction for Extended
  Virasoro Algebras, \np  {\bf B304} (1988) 348, 371
\bibitem{bpz}
  A.A. Belavin, A.M. Polyakov and A.B. Zamolodchikov, Infinite
  Conformal Symmetry in Two-Dimensional Quantum Field Theory, \np
  {\bf B241} (1984) 333
\bibitem{work}
  C. Ahn, work in progress
\end{thebibliography}
\end{document}